\documentstyle[prd,aps,epsfig]{revtex}

\def\lsim{\mathrel{\rlap{\lower4pt\hbox{\hskip1pt$\sim$}}
    \raise1pt\hbox{$<$}}}         
\def\gsim{\mathrel{\rlap{\lower4pt\hbox{\hskip1pt$\sim$}}
    \raise1pt\hbox{$>$}}}         

\begin{document}
 
 

\title{{\rm\small \hfill MPI-PhT/97-62}\\
Limits on R-parity violation from cosmic ray antiprotons}
 
\author{Edward A. Baltz\footnote{E-mail address: eabaltz@wharton.berkeley.edu}}
\address{Department of Physics, University of California,\\
 601 Campbell Hall, Berkeley, CA 94720-3411}
\author{Paolo Gondolo\footnote{E-mail address: gondolo@mppmu.mpg.de}}
\address{Center for Particle Astrophysics, University of California,\\
301 Le Conte Hall, Berkeley, CA 94720-7304\\
{\rm and}\\
Max-Planck-Institut f\"{u}r Physik, Werner-Heisenberg-Institut,
\footnote{Present address.}\\
F\"ohringer Ring 6, 80805 M\"unchen, Germany.}

 
\maketitle

\vspace{.5in}
 
\begin{abstract}  
We constrain the hadronic R-parity violating couplings in extensions to the
minimal supersymmetric standard model.  These interactions violate baryon and
lepton number, and allow the lightest superpartner (LSP) to decay into standard
model particles.  The observed flux of cosmic ray antiprotons places a strong
bound on the lifetime of the LSP in models where the lifetime is longer than
the age of the universe.  We exclude $10^{-18}\lsim |\lambda''|\lsim
10^{-15}$ and $2\times 10^{-18}\lsim |\lambda'|\lsim 2\times 10^{-15}$
except in the case of a top quark, where we can only exclude $4\times
10^{-19}\lsim |\lambda'|\lsim 4\times 10^{-16}$.
\end{abstract}

 
\twocolumn
\narrowtext
 
\section{Introduction} \label{sec:intro}

In supersymmetric models, gauge invariance allows interactions that violate
lepton- or baryon-number conservation. To avoid them, a discrete symmetry
called R-parity is often imposed. The R-parity of a particle is defined as
$R=(-1)^{2S+L+3B}$, where $S$ is its spin, and $L$ and $B$ are its lepton and
baryon numbers. $R$ is $+1$ for all standard model particles and $-1$ for all
superparticles. Alternatively, it is an interesting exercise to allow R-parity
violating couplings and constrain them by experimental observations or
cosmological considerations. With R-parity violation, the lightest superpartner
decays into standard model particles. We will assume that the lightest
superpartner is the neutralino.

Antiprotons are a rare component of cosmic rays, most probably produced in
spallation reactions in the interstellar medium. The measured antiproton flux
limits any additional antiproton production in the galaxy.  We find a powerful
bound on the lifetime of the lightest superpartner from the observed flux of
cosmic ray antiprotons.  This allows us to bound the R-parity violating
interactions that appear in extensions to the minimal supersymmetric standard
model.

We find the relic density of the lightest superpartner (LSP) using the standard
cosmological model.  We then allow these particles to decay via the hadronic
R-parity violating operators.  We employ simple models for the production rate
and the propagation of the antiprotons.  We can limit the R-parity violating
couplings with observations of the antiproton spectrum in cosmic rays.

Our results strictly apply when the lifetime of the neutralino is longer than
the age of the universe.  However, there are other cosmological limits based on
nucleosynthesis in the early universe, recent structure formation, diffuse
photon and neutrino fluxes, and spectral distortions in the cosmic microwave
background that cover a large fraction of neutralino lifetimes shorter than the
age of the universe, and may well completely close the gap between this result
and laboratory constraints.

\section{Supersymmetric model} \label{sec:susy}

The supersymmetric model we consider is a minimal extension of the standard
model with explicit R-parity breaking, defined by the superpotential
\begin{eqnarray}
  W  =
  - h_e^{ij} E^c_i L_j H_1
  - h_d^{ij} D^c_i Q_j H_1
  + h_u^{ij} U^c_i Q_j H_2 \nonumber \\
  - \mu H_1 H_2
  + \lambda'_{ijk} L_i Q_j D^c_k
  + \lambda''_{ijk} U^c_i D^c_j D^c_k .
\end{eqnarray}
$i,j,k$ are generation indices; $L_i$ and $Q_i$ are the SU(2)-doublet lepton
and quark superfields; $E^c_i$, $U^c_i$ and $D^c_i$ are the singlet
superfields; and $H_1$ and $H_2$ are the doublet Higgs superfields. $h_e^{ij}$,
$h_d^{ij}$ and $h_u^{ij}$ are Yukawa matrices, and $\mu$ is a free parameter
\cite{susy}.

The last two terms in the superpotential violate lepton- and baryon-number
conservation explicitly. For three generations, there are 27 $\lambda'$
couplings and 9 $\lambda''$ couplings, since $\lambda''_{ijk}$ is antisymmetric
under exchange of the last two indices.

Supersymmetry is broken softly by a scalar potential that in general contains
many unmeasured parameters. For our purposes, a simple parameterization
suffices. We assign a common mass scale $m_0$ to the sleptons and squarks at
the electroweak scale, we relate the gaugino mass parameters $M_1, M_2, M_3$ by
means of GUT relations, and we keep the trilinear soft terms $A_t$ and $A_b$
only for stop and sbottom. Electroweak symmetry is broken at tree level by
assigning vacuum expectation values $v_1$ and $v_2$ to the neutral scalar
components of $H_1$ and $H_2$. The ratio $\tan\beta=v_2/v_1$ and the
pseudoscalar Higgs boson mass $m_A$ remain as free parameters. The total number
of parameters is therefore seven. Further details on the class of models we
consider are given by Bergstr\"{o}m and Gondolo \cite{bg}.

We use the scans of supersymmetric parameter space in Bergstr\"{o}m,
Edsj\"{o}, and Gondolo \cite{beg}.  They consist of $\sim 10^4$ models that
satisfy the experimental constraints reported by the Particle Data Group
\cite{pdg}, plus the bounds on the Higgs sector and chargino masses from LEP
(we update to LEP2W~\cite{lep2w}) and the constraints on the $b\to s\gamma$
rate from CLEO \cite{susylimits}.  The scans span values of $|\mu|$ and
$|M_2|$ up to 5 TeV, of $m_0$ up to 3 TeV, of $|A_t|$ and $|A_b|$ up to
$3m_0$, of $m_A$ up to 1 TeV, and of $\tan\beta$ between 1 and 50. This range
extends even beyond the parameter region usually considered as plausible.

\section{Composition of the Galactic Halo}

We do not assume that the dark matter is supersymmetric in nature.  Given a
supersymmetric model, the relic density of neutralinos is fixed and calculable
in the standard cosmological model. We use the relic density computed by
Edsj\"{o} and Gondolo \cite{eg}, who include all neutralino self-annihilation
diagrams, resonance and threshold effects, and chargino-neutralino
coannihilations.  In every model that is cosmologically allowed, the lightest
neutralino is a cold relic from the Big Bang.  Being a cold relic, we expect
that it is attracted into the gravitational potentials of galaxies, and is at
present a component of the cold dark halo.

We use the simple estimate that the dark matter density of the galactic halo at
the position of the solar system is 0.3 GeV cm$^{-3}$.  Neutralinos may make up
the entire halo or only some fraction $f$.  This fraction
$f\ge\Omega_\chi/\Omega_{\rm DM}$, where $\Omega_A$ is the cosmological density
of species $A$ measured in units of the critical density.

We now need an estimate of $\Omega_{\rm DM}$.  A common prescription
\cite{susydm} is to take $\Omega_{\rm DM}=\Omega_\chi$ if $\Omega_\chi h^2\ge
0.025$, which is the value indicated by dynamical mass measurements of the
amount of dark matter in galactic halos.  If $\Omega_\chi h^2\le 0.025$, this
prescription sets $\Omega_{\rm DM}h^2=0.025$.  This procedure is not
conservative enough for our purposes because there may be more than one cold
dark matter component in the galactic halo.  Therefore, we use the constraint
that $\Omega_{\rm DM} h^2<1$, based on the age of the universe \cite{susydm},
to give a lower bound on $f=\Omega_\chi h^2$.  In this way we obtain a
conservative limit on the R-parity violating couplings.  This may be overly
conservative, and indeed our final results would improve by approximately a
factor of five if we used the less conservative fraction.

\section{Decay rate} \label{sec:decay}

We consider neutralino decay through both the lepton-number violating term
$\lambda'LQD^c$ and the baryon-number violating term $\lambda''U^cD^cD^c$ in
the superpotential.  The coupling $\lambda'_{ijk}$ gives rise to the decays
$\chi\to \nu_i d_j \bar{d}_k$, $\chi \to e_i^- u_j \bar{d}_k$ and their charge
conjugates. The coupling $\lambda''_{ijk}$ gives rise to the decays $\chi\to
u_i d_j d_k$ ($j\ne k$) and their charge conjugates.

The differential decay rates for these processes are calculated in Ref.
\cite{rates}.  Here we quote the decay rate for $\chi\to uds$ in the limit
where the squarks are heavy and degenerate,
\begin{eqnarray}
&&\Gamma(\chi\to uds)=\frac{3}{32\pi^3}m_\chi|\lambda''_{112}|^2g^2
\left(\frac{m_\chi}{m_{\tilde{q}}}\right)^4\times \\
&&\left[2\tan^2\theta_W|N_{\chi 1}|^2 +\frac{m_u^2|N_{\chi 4}|^2}
{2m_W^2\sin^2\beta}+\frac{(m_d^2+m_s^2)|N_{\chi3}|^2}{2m_W^2\cos^2\beta}
\right] \nonumber,
\end{eqnarray}
where $g$ is the $SU(2)_L$ coupling constant, $m_{\tilde{q}}$ is the degenerate
squark mass, and $N_{\chi i}$ is the projection of the lightest neutralino on
the interaction eigenstates $\tilde{b}$, $\tilde{w}^0$, $\tilde{h}_1$, and
$\tilde{h}_2$.  When the gaugino fraction approaches unity, we see that the
decay rate is
\begin{equation}
\Gamma = \frac{3}{16\pi^3}m_\chi|\lambda''_{112}|^2g^2\tan^2\theta_W\left(
\frac{m_\chi}{m_{\tilde{q}}}\right)^4.
\end{equation}
When the gaugino fraction approaches zero, there is an additional suppression
from the light quark masses,
\begin{eqnarray}
\Gamma = \frac{3}{128\pi^3}m_\chi|\lambda''_{112}|^2&&g^2\left(
\frac{m_\chi}{m_{\tilde{q}}}\right)^4\times \\
&&\left[\frac{m_u^2}{m_W^2\sin^2\beta}+
\frac{m_d^2+m_s^2}{m_W^2\cos^2\beta}\right] \nonumber.
\end{eqnarray}

\section{Antiproton production} \label{sec:jets}

Antiprotons are produced in the jets formed by the final state quarks.  We are
interested in the energy spectrum of antiprotons $dN_{\overline{p}}/dE$ at
energies of a few GeV.

It is quite difficult to calculate the spectrum $dN_{\overline{p}}/dE$,
especially at low antiproton energies. There are basically two theoretical
frameworks available: a perturbative QCD approach and the Lund Monte-Carlo.
In fig.~\ref{fig:pbar} we compare the theoretical curves with the measured
proton spectra per quark jet in $e^+e^-$ annihilations at various
jet energies (14.5 GeV~\cite{TPC}, 29 GeV~\cite{TOPAZ}, and 45.6
GeV~\cite{LEP}) and also show the theoretical predictions at 1 TeV, one of
the highest energies we need.

In the perturbative QCD approach, Dokshitzer et al. \cite{doksh} provide a
formalism for calculating hadron spectra from parton spectra in the modified
double-logarithmic approximation (MLLA). The full MLLA formula for inclusive
particle production is too difficult to evaluate, and no numerical expression
is available.  It has been shown that the simpler limiting spectrum reproduces
the total charged particle spectrum quite well \cite{softqcd}.  However, as
can be seen in fig.~\ref{fig:pbar}, neither the limiting spectrum (long-dashed
curves~\cite{doksh}) nor the low energy limit of the full MLLA spectrum
(short-dashed curves~\cite{softqcd}) fit data for protons in a satisfactory
manner.  As a theoretical alternative to the MLLA formalism, the Lund Monte
Carlo~\cite{lund} has been used by several authors, most recently by Bottino
et al.~\cite{bottino}. However, it also does not match the data for protons
satisfactorily (histograms in fig.~\ref{fig:pbar}).

Inspecting the experimental data and the results of the Lund Monte Carlo, we
observe that the (anti)proton production rate in the high-energy tail of the
spectrum increases with jet energy. This is verified in the data for charged
particle production in jets and is a generic prediction of the MLLA formalism
\cite{doksh,softqcd}.

Based on these arguments, we decide to take the value
$dN_{\overline{p}}/dE=0.05\;{\rm antiprotons/jet/GeV}$ at the momentum
$p=1.573$ GeV in which we want the antiproton spectra (see
sect.~\ref{sec:res}).  This underestimates the production rate of antiprotons,
and gives a conservative bound on the R-parity violating couplings.

The number of jets per decay $N_{\rm jets}$ is equal to the number of final
state quarks, except when there is a top quark.  The top decays before it can
hadronize.  We approximate the process $t\to W^+b$ using the fact that the low
energy proton spectrum is insensitive to the jet energy.  The bottom quark
forms a jet.  In addition, we take the hadronic branching fraction of the $W$,
$B(W\to {\rm hadrons}) = 0.679$, as the probability to form two additional jets
in the decay $W^+\to u\overline{d}$ and its charge conjugate.  The final values
for $N_{\rm jets}$ are 3 for a $U^cD^cD^c$ decay without top, 4.358 for a
$U^cD^cD^c$ with top, 2 for an $LQD^c$ without top and 3.358 for an $LQD^c$
with top.

We now obtain the volume production rate of antiprotons by multiplying the flat
antiproton spectrum by the number density of neutralinos $\rho_\chi/m_\chi$ and
the number of jets per decay $N_{\rm jets}$, and dividing by the lifetime
$\tau_\chi=1/\Gamma$,
\begin{equation}
\frac{dQ_{\overline{p}}}{dE}=\frac{dN_{\overline{p}}}{dE}
\frac{\rho_\chi}{m_\chi}
\frac{N_{\rm jets}}{\tau_\chi}.
\end{equation}

\section{Antiproton propagation in the galaxy} \label{sec:diff}

We use the diffusion model of Webber {\em et al.} \cite{crprop} as implemented
by Chardonnet {\em et al.} \cite{taillet} to describe diffusion of antiprotons
through the galaxy.  The flux of antiprotons at the outer solar system is
approximated by multiplying the source spectrum by the antiproton velocity and
an appropriate diffusion time.  The diffusion constant in the galaxy depends on
the rigidity $p$ of the particle and can be approximated by the following
expression.
\begin{equation}
K=6\times 10^{27}\left(1+\frac{p}{3{Z\rm GV}}\right)^{0.6}
{\rm cm}^2\;{\rm sec}^{-1}.
\end{equation}
The region of turbulent magnetic fields responsible for diffusion can be
approximated as a cylinder about 20kpc in radius and 2kpc high.  This gives a
diffusion time for antiprotons of
\begin{equation}
t_d=6\times 10^{15}\left(1+\frac{p}{3{\rm GV}}\right)^{-0.6}\;{\rm sec}.
\end{equation}
The flux at the outer solar system beyond the influence of the solar wind is
now given by
\begin{equation}
\Phi_{\overline{p}}=\frac{1}{4\pi}
\frac{dQ_{\overline{p}}}{dE}v_{\overline{p}}t_d .
\end{equation}
This flux is in units of antiprotons s$^{-1}$ cm$^{-2}$ GeV$^{-1}$ sr$^{-1}$.

The low energy spectrum is affected by the solar wind.  We use a simple model
to account for this \cite{perko}.  The observed antiproton energy $E$ and
momentum $k$ are given in terms of the initial energy outside the solar system
$E_{\overline{p}}$.
\begin{eqnarray}
E_{\overline{p}} & = & k_c\ln\frac{k+E}{k_c+E_c}+E_c+\Delta E ,
\hspace{0.2in} k<k_c,\\
\nonumber
E_{\overline{p}} & = & E + \Delta E, \hspace{0.2in} k \ge k_c.
\end{eqnarray}
We adopt values of the critical momentum $k_c$ and energy shift $\Delta E$ that
correspond to the period of minimum activity in the 11 year solar cycle,
$k_c=$1.105 GeV and $\Delta E$=495 MeV.  This gives the highest inner solar
system flux, thus the conservative bound on the couplings.

The solar modulated flux is now given by
\begin{equation}
\Phi_{\overline{p}}(E)=\frac{E^2-m^2_p}{E_{\overline{p}}^2-m^2_p}
\Phi_{\overline{p}}(E_{\overline{p}}) .
\end{equation}

\section{Results} \label{sec:res}

We proceed as follows. We choose a data point for the $\overline{p}$ flux at
the earth \cite{crdata}. These data have large error bars, but will improve
with the next generation of experiments to be done in space \cite{crfuture}.
According to the data from BESS, at an energy of 1.4 GeV the antiproton flux at
the top of the atmosphere is 6.4 $\times 10^{-7} [{\rm cm}^2\,{\rm
s\,sr\,GeV}]^{-1}$.  We denote this flux as $\Phi_{\rm obs}(E)$.  From the
values of the observed energy $E$ and momentum $k$, we determine the
unmodulated $\overline{p}$ energy $E_{\overline{p}}$.  This is the energy at
which we evaluate the diffusion time $t_d$ and the velocity $v$.

We now insist that the flux of antiprotons from neutralino decays be less than
the observed flux.  Including corrections due to solar modulation, the
statement is
\begin{equation}
\frac{v_{\overline{p}}t_d}{4\pi}\frac{\rho_\chi}{m_\chi}
\frac{N_{\rm jets}}{\tau_\chi}\frac{dN_{\overline{p}}}{dE}
\frac{E^2-m^2_{\overline{p}}}{E^2_{\overline{p}}-m_{\overline{p}}}
<\Phi_{\rm obs}(E).
\end{equation}

We choose not to make a correction for the spallation production of
antiprotons.  The data are consistent with the picture that all cosmic ray
antiprotons are made in spallation reactions, but there are a lot of
uncertainties.  There is no good model for the amount we should subtract, and
we have chosen to ignore it.  We thus obtain a conservative bound on the
couplings.

Using the data we have chosen, we obtain a model independent bound on the
lifetime of a hadronically decaying relic with mass $m_\chi \gsim 10$ GeV
\begin{equation}
  \tau_\chi \gsim \tau_\chi^{\bar{p}} =
  7.9\times 10^{28}\,{\rm s}\, \frac{N_{\rm jets}
    f_{\chi}}{m_\chi/{\rm GeV}},
\end{equation}
where we choose the mass fraction of neutralinos in the
galactic dark halo as $f_\chi=\Omega_\chi h^2$.  Since we
assume that the neutralino is in the galactic halo at
present, this bound applies for lifetimes longer than the
age of the Universe $t_0$. In other words, we can exclude
lifetimes in the range 
$t_0 \le \tau_\chi \le \tau_\chi^{\bar{p}}$
provided $t_0 \le \tau_\chi^{\bar{p}}$.

For these long lifetimes, our bounds are in general better than those from
diffuse photons \cite{diffgamma}, $\tau_\chi \gsim 7 \times 10^{25} {\rm s}\,
N_{\rm jets} \Omega_\chi h^2$, and diffuse neutrinos \cite{wumps}, $\tau_\chi
\gsim 10^{25}{\rm s}\,(m_\chi/{\rm GeV})^{1.7}f_\chi$, except for some
supersymmetric models with neutralino masses in the TeV range.

We can limit each element of the two coupling matrices
$\lambda'$ and $\lambda''$ separately, by choosing the
coupling such that all of the measured $\overline{p}$ flux
comes from the neutralino decay via a single channel. Our
excluded range of lifetimes then translates into the
following {\it excluded} range of $\lambda$
\begin{equation}
\lambda_{\bar{p}} \le |\lambda| \le \lambda_0,
\end{equation}
where
\begin{equation}
\lambda_{\bar{p}} \simeq 
3.48\times 10^{-15}\sqrt{\frac{m_\chi\tilde{\tau}}{N_{\rm jets}f_{\chi}}}
\end{equation}
is the upper limit on $\lambda$ coming from the antiproton
flux and
\begin{equation}
\lambda_0 \simeq 1.8\times 10^{-9}\sqrt{\tilde{\tau}}
\end{equation}
is the value of $\lambda$ for which the neutralino lifetime
equals the age of the universe, which sets the validity of our analysis.
In the previous formulas,  $m_\chi$ is in GeV,
$\tilde{\tau}=|\lambda|^2\tau_{\chi}$ is in seconds, and the
age of the universe is taken to be 
$10^{10}$ yr.

For virtually all models we
find that $\lambda_{\bar{p}}$ is smaller than $\lambda_0$ by
at least three orders of magnitude.
Conservatively taking $N_{\rm jets}=2$ we find for the excluded range that
\begin{equation}
\frac{\lambda_{0}}{\lambda_{\bar{p}}}=7.3\times
10^5\sqrt{\frac{\Omega_\chi h^2}{m_\chi/{\rm GeV}}}\gsim 600.
\end{equation}
If we would further demand that the neutralino be
cosmologically interesting, namely $\Omega_\chi h^2 >
0.025$, we would expand the excluded range to
$\lambda_{0}/\lambda_{\bar{p}} \gsim 4000$.

We have plotted the upper bound on $\lambda''_{112}$ as a function of the
neutralino mass for 10$^4$ supersymmetric models in figure~\ref{fig:llmx}.
The results are similar for all generations in the R-parity violating
couplings, except for $\lambda''_{3jk}$ which shows the top quark threshold
(see figure~ \ref{fig:lltop}).  When the neutralino is lighter than the top
quark, there is no bound on $\lambda''_{3jk}$ at tree level. A bound can
however be found at the one loop level \cite{toploop}.

In our sample of supersymmetric models, we find the absolute upper limits
\begin{eqnarray}
|\lambda'_{ijk}|_{\bar{p}} && \lsim 2\times 10^{-18} \qquad (i,j,k\ne3) \\
|\lambda'_{3jk}|_{\bar{p}} && \lsim 1\times 10^{-18} \qquad (j,k\ne3) \\
|\lambda'_{i3k}|_{\bar{p}} && \lsim 4\times 10^{-19} \qquad (k\ne3) \\
|\lambda'_{ij3}|_{\bar{p}} && \lsim 3\times 10^{-19} \\
|\lambda''_{ijk}|_{\bar{p}} && \lsim 1\times 10^{-18} \qquad (i,k\ne3) \\
|\lambda''_{ij3}|_{\bar{p}} && \lsim 1\times 10^{-19} \qquad (i\ne3).
\end{eqnarray}
The apparent absence of an absolute upper bound on $\lambda''_{3ij}$ is due to
our neglect of one-loop diagrams.  These upper limits depend on the neutralino
mass, and improve considerably with higher masses.  We find the following power
law fits
\begin{eqnarray}
|\lambda'_{ijk}|_{\bar{p}} && \lsim 6\times 10^{-22}(m_\chi/{\rm TeV})^{-3.1}
\qquad (i,j,k\ne3) \\
|\lambda'_{3jk}|_{\bar{p}} && \lsim 3\times 10^{-22}(m_\chi/{\rm TeV})^{-3.1}
\qquad (j,k\ne3) \\
|\lambda'_{i3k}|_{\bar{p}} && \lsim 8\times 10^{-23}(m_\chi/{\rm TeV})^{-3.2}
\qquad (k\ne3) \\
|\lambda'_{ij3}|_{\bar{p}} && \lsim 1\times 10^{-22}(m_\chi/{\rm TeV})^{-3.0} \\
|\lambda''_{ijk}|_{\bar{p}} && \lsim 3\times 10^{-22}(m_\chi/{\rm TeV})^{-2.7}
\qquad (i,k\ne3) \\
|\lambda''_{ij3}|_{\bar{p}} && \lsim 3\times 10^{-23}(m_\chi/{\rm TeV})^{-2.7}
\qquad (i\ne3).
\end{eqnarray}

For comparison, the most stringent laboratory constraints,
those from proton decay~\cite{hinchliffe}, bound not
$|\lambda'|$ and $|\lambda''|$ separately but their product
to $|\lambda' \lambda''| \lsim 10^{-24} (m_0/{\rm TeV})^2$,
where $m_0$ is the sfermion mass scale.  These bounds are six
orders of magnitude weaker than ours. On the other hand,
because our analysis only applies to lifetimes longer than
the age of the Universe, our excluded region spans only 3
orders of magnitude. As mentioned in the introduction, the
remaining gap of 3 orders of magnitude may be filled by
other cosmological considerations.

We finally comment on the dependence of our bounds on model parameters.  In
figure~\ref{fig:llmsq} we plot the dependence on the squark mass scale
$m_0$. As expected, the bound is better for lighter squarks. Regarding the
dependence on neutralino composition, in general the bounds are better in the
gaugino region.  In the higgsino region, the bounds improve with heavier
generations as the Yukawa couplings become larger. The dependence on other
model parameters is negligible in comparison.

\section{Conclusions}

{}From observations of cosmic ray antiprotons we obtain strong constraints on
each of the hadronic R-parity violating couplings $|\lambda'|$ and
$|\lambda''|$ that appear in extensions of the minimal supersymmetric
model. Our analysis applies strictly in cases where the neutralino lifetime is
longer than the age of the universe, and enables us to exclude a range
$\lambda_{\bar{p}} \le |\lambda| \le \lambda_{0}$, where $\lambda_{\bar{p}}$ is
the upper limit from the antiproton flux and $\lambda_0$ is the validity
boundary coming from the age of the universe.  Using a large selection of model
parameters, we find absolute upper limits $|\lambda'_{ijk}|_{\bar{p}} \lsim
2\times 10^{-18}$ and $|\lambda''_{ijk}|_{\bar{p}} \lsim 10^{-18}$ when there
is no top quark in the final state, and $|\lambda'_{i3k}|_{\bar{p}} \lsim
4\times 10^{-19}$ when there is. In virtually all models, we find that
$\lambda_0 \gsim 10^3 \lambda_{\bar{p}}$.  Our bounds lie many orders of
magnitude below the laboratory constraints coming from proton decay, and apply
to each R-parity violating coupling separately.

\section*{Acknowledgments}

We thank J. Silk for useful discussions and comments on the manuscript.  P.G.
thanks him also for the friendly hospitality and encouraging support at the
Center for Particle Astrophysics. P.G. thanks S. Lupia for useful discussions
and programs on the theoretical proton spectra.  This research was supported
in part by NASA grant 1-443839-23254-2 and DOE grant FG03-84ER40161.

\setlength{\baselineskip}{.5\baselineskip}
 
\onecolumn
\widetext

 
\clearpage
 
\setlength{\parindent}{0cm}
 
\begin{figure}[h]
\caption{Antiproton spectra in quark jets 
  from theoretical predictions and $e^+e^-$ annihilation data.  Short- and
  long-dashed curves are the low-energy and the limiting MLLA spectra,
  histograms are obtained with the Lund Monte-Carlo. Data points are from TPC
  (14.5 GeV), TOPAZ (29 GeV), ALEPH (41.6 GeV, full squares), DELPHI (41.6 GeV,
  diamonds) and OPAL (41.6 GeV, open squares). The cross indicates our choice
  of $dN_{\overline{p}}/dE$ at $p=1.573$ GeV.  }
\label{fig:pbar}
\end{figure}

\begin{figure}[h]
  \caption{Bounds on $\lambda''_{112}$ from cosmic ray antiprotons as a
    function of neutralino mass for 10$^4$ SUSY models.  The scatter plots for
    $\lambda''_{ijk}$ are similar when $i\ne 3$.  The scatter plots for
    $\lambda'_{ijk}$ are also similar.}
  \label{fig:llmx}
\end{figure}
 
\begin{figure}[h]
  \caption{Bounds on $\lambda''_{312}$ from cosmic ray antiprotons as a
    function of the neutralino mass for 10$^4$ SUSY models.  The threshold
    effect is clearly seen.}
  \label{fig:lltop}
\end{figure}

\begin{figure}[h]
  \caption{Bounds on $\lambda''_{112}$ from cosmic ray antiprotons as a
    function of the squark mass scale for 10$^4$ SUSY models.}
  \label{fig:llmsq}
\end{figure}

 
\clearpage Fig.~\ref{fig:pbar}, E.~A.~Baltz and P.~Gondolo, Phys.\ Rev.\ 
{\bfseries D}
\\[5ex] \centerline{\epsfig{file=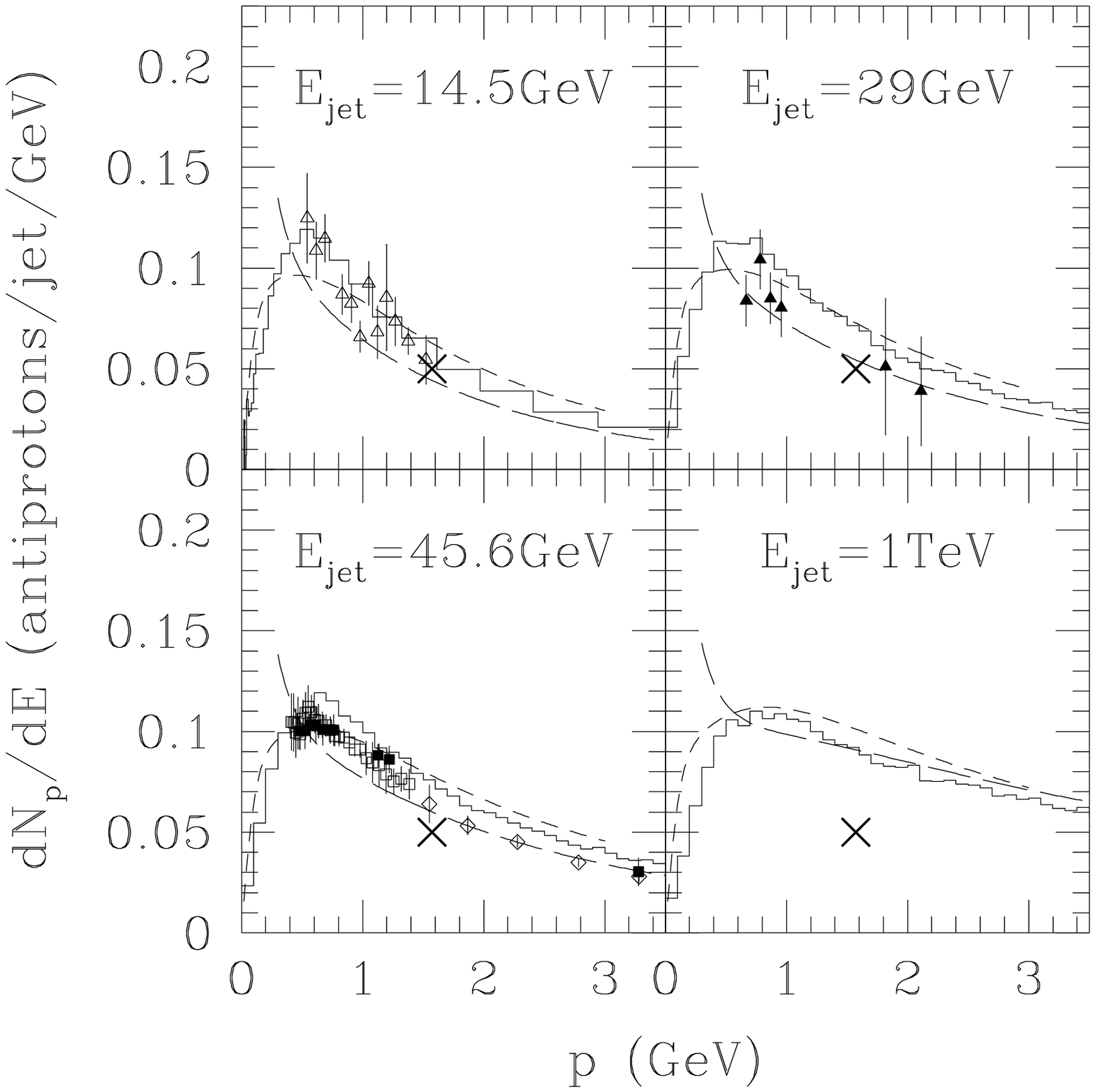}}

\clearpage Fig.~\ref{fig:llmx} E.~A.~Baltz and P.~Gondolo, Phys.\ Rev. \ 
{\bfseries D}
\\[5ex] \centerline{\epsfig{file=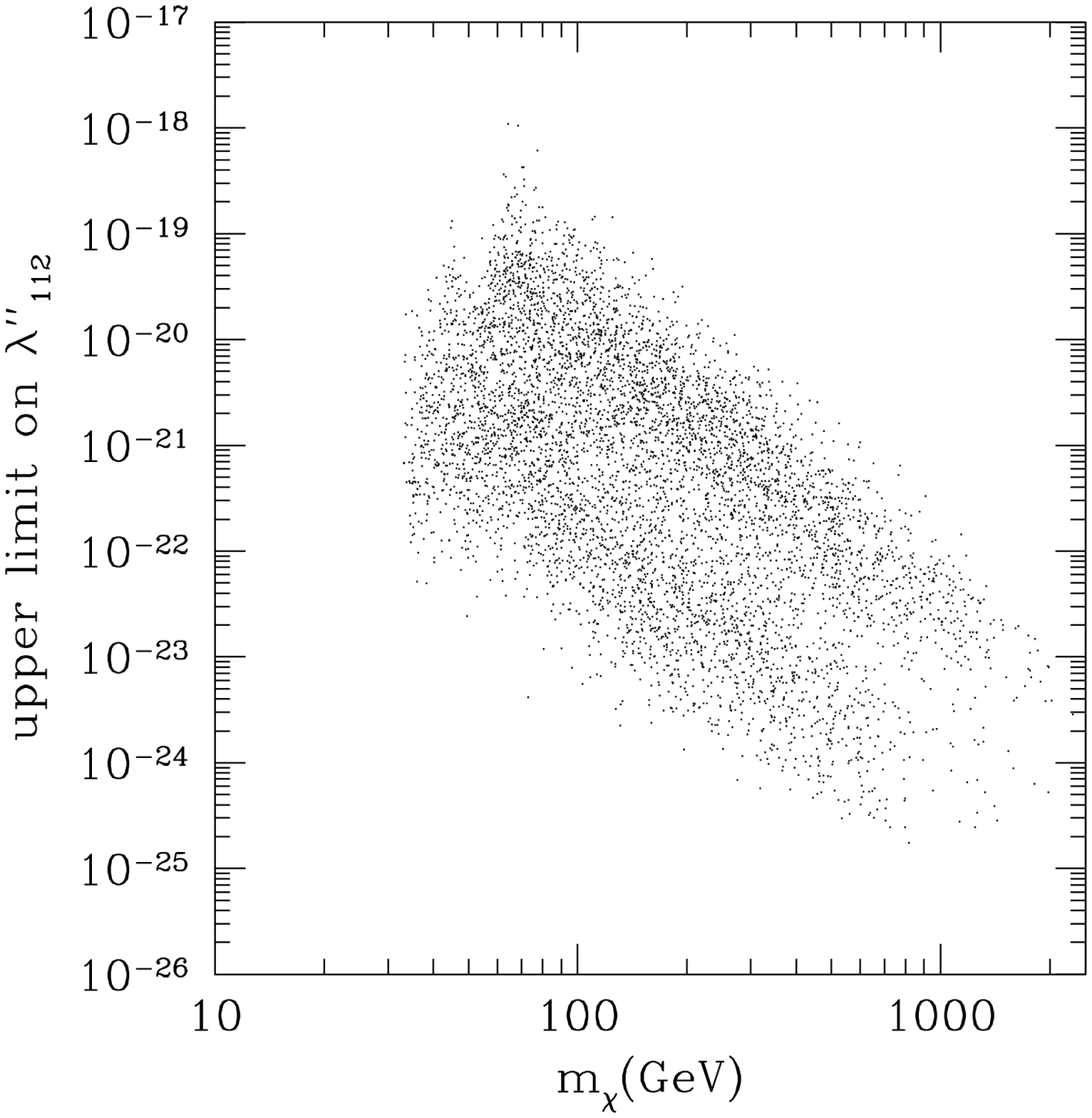}}

\clearpage Fig.~\ref{fig:lltop} E.~A.~Baltz and P.~Gondolo, Phys.\ Rev. \ 
{\bfseries D}
\\[5ex] \centerline{\epsfig{file=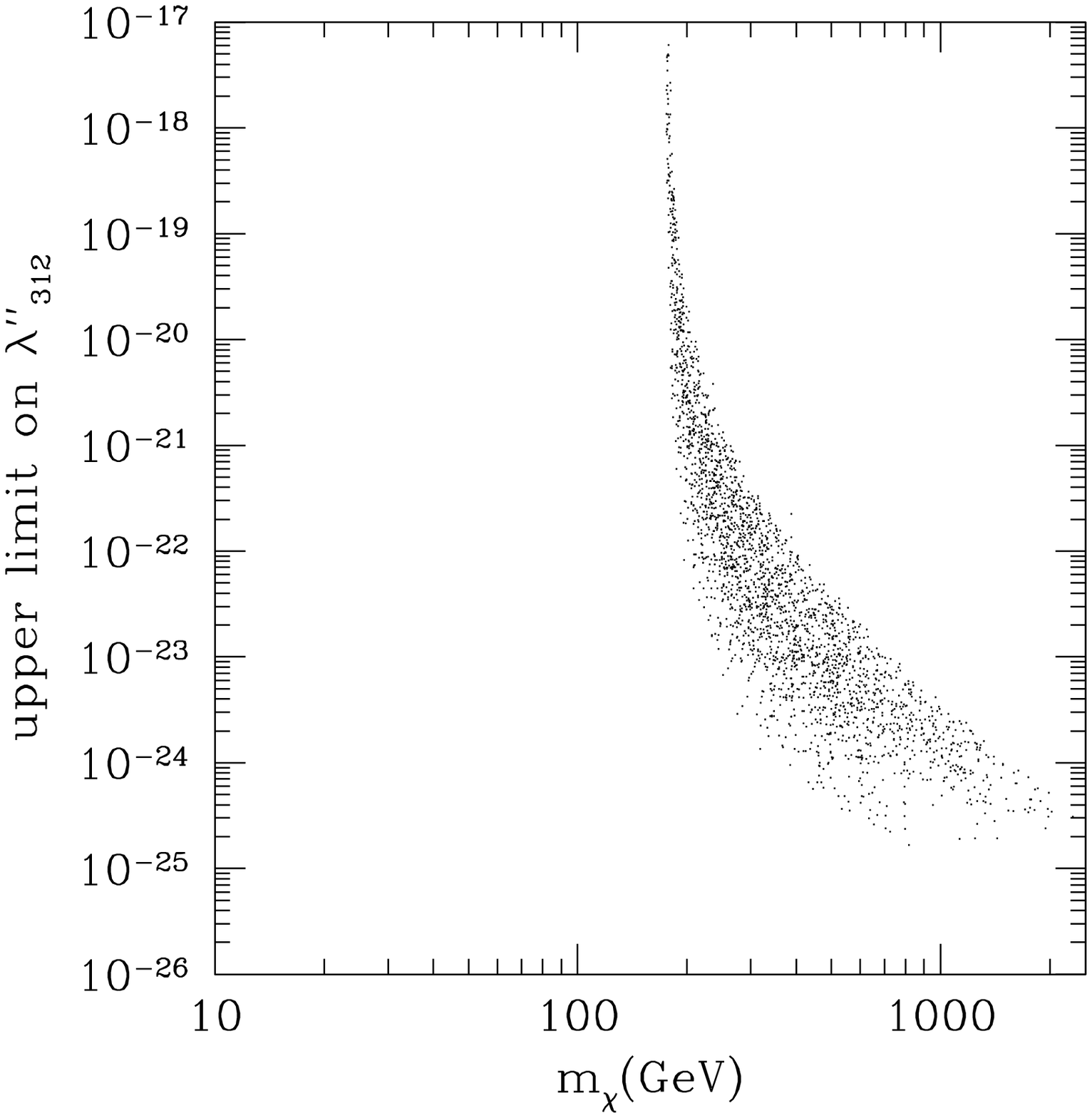}}

\clearpage Fig.~\ref{fig:llmsq} E.~A.~Baltz and P.~Gondolo, Phys.\ Rev. \ 
{\bfseries D}
\\[5ex] \centerline{\epsfig{file=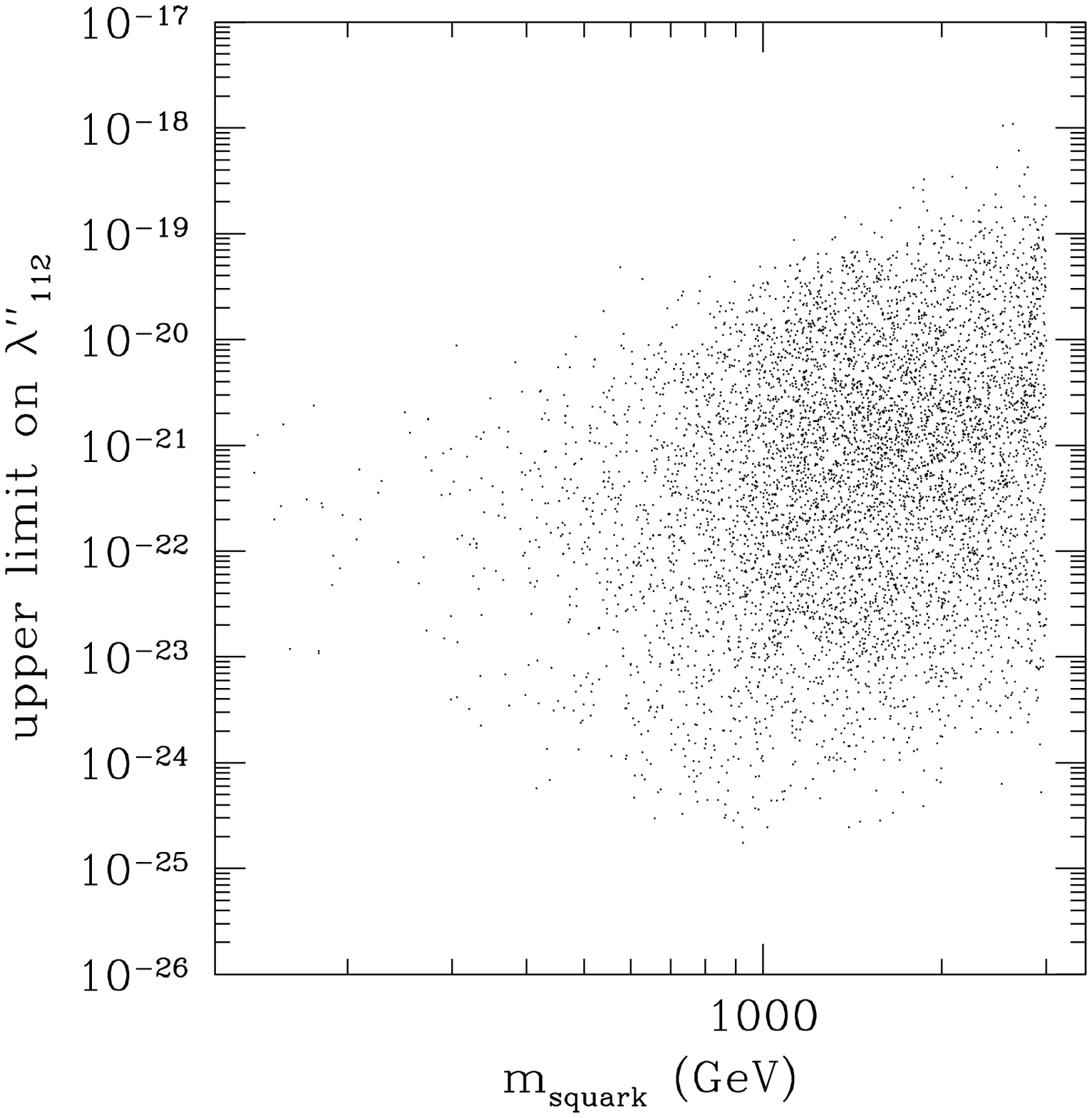}}

\end{document}